\documentclass[12pt,letter]{article}

\usepackage{graphicx, epsfig, color}
\usepackage{amssymb}
\def\eslt{E_T^{\rm miss}}
\def\delew{\Delta_{\rm EW}}
\def\delhs{\Delta_{\rm HS}}
\def\delbg{\Delta_{\rm BG}}
\def\to{\rightarrow}

\def\bi{\begin{itemize}}
\def\ei{\end{itemize}}

\def\tb{\tilde b}

\def\tst{\tilde t}

\def\tg{\tilde g}

\def\tq{\tilde q}
\def\tw{\widetilde W}
\def\tz{\widetilde Z}
\def\be{\begin{equation}}  
\def\ee{\end{equation}}  
\def\bea{\begin{eqnarray}}  
\def\eea{\end{eqnarray}}  
\def\beas{\begin{eqnarray*}}  
\def\eeas{\end{eqnarray*}}  
\newcommand\prd[3]{{\it Phys.\ Rev.\ }{\bf D #1} (#2) #3}

\newcommand\prl[3]{{\it Phys.\ Rev.\ Lett.\ }{\bf #1} (#2) #3}
\newcommand\plb[3]{{\it Phys.\ Lett.\ }{\bf B #1} (#2) #3}
\newcommand\jhep[3]{{\it J. High Energy Phys.\ }{\bf #1} (#2) #3}

\newcommand\npb[3]{{\it Nucl.\ Phys.\ }{\bf B #1} (#2) #3}

\newcommand\ptp[3]{{\it Prog.\ Theor.\ Phys.\ }{\bf #1} (#2) #3}

\newcommand{\hepph}[1]{hep-ph/#1}

\newcommand\ppnp[3]{{\it Prog.\ Part.\ Nucl.\ Phys.}{\bf  #1} (#2) #3}

\begin{document}
\begin{titlepage}
\begin{flushright}
UH-511-1231-14
\end{flushright}

\vspace{0.5cm}
\begin{center}
\renewcommand{\thefootnote}{\fnsymbol{footnote}}
{\Large \bf Supersymmetry, Naturalness, and Light Higgsinos\footnote[1]{Invited contribution
to the Volume Commemorating C.~V.~Raman's 125th Birth Anniversary}
}\\ 
\vspace{0.5cm} 
{\large Azar Mustafayev\footnote[3]{Email: azar@phys.hawaii.edu }
and Xerxes Tata\footnote[2]{Email: tata@phys.hawaii.edu } }
  
\vspace{0.5cm} \renewcommand{\thefootnote}{\arabic{footnote}}

{\it 
Dept. of Physics and Astronomy,
University of Hawaii, Honolulu, HI 96822, USA \\
}

\end{center}

\vspace{0.5cm}
\begin{abstract}
\noindent 
We compare and contrast three different sensitivity measures,
$\delew^{-1}$, $\delhs^{-1}$ and $\delbg^{-1}$ that have been used in
discussions of fine-tuning. We argue that though not a fine-tuning
measure, $\delew$, which is essentially determined by the particle
spectrum, is important because $\delew^{-1}$ quantifies the {\em
minimum} fine-tuning present in any theory with a specified spectrum. We
emphasize the critical role of incorporating correlations between
various model parameters in discussions of fine-tuning. We provide toy
examples to show that if we can find high scale theories with specific
correlations amongst parameters, the value of the traditional
fine-tuning measure $\delbg^{-1}$ (which differs significantly from
$\delhs^{-1}$ only when these correlations are important) would be close
to $\delew^{-1}$. We then set up the radiatively driven natural SUSY
framework that we advocate for phenomenological analyses of natural
models of supersymmetry, and review the implications of naturalness for
LHC and ILC searches for SUSY as well as for searches for SUSY dark
matter.

\vfill
\noindent Keywords: supersymmetry, naturalness, higgsino signatures, Large
          Hadron Collider,\\  Linear Collider 

\noindent PACS numbers: 14.80Lv, 12.60Jv

\end{abstract}

\end{titlepage}

\section{Introduction} 
\label{sec:intro}

It is common knowledge that effective theories valid above some
distance scale provide an excellent description of phenomena down to that 
scale. Hydrodynamics does not require us to even know about the
existence of atoms, and applies at distance scales much larger than
the size of atoms.
Likewise, an understanding of the atoms and molecules does not require
knowledge of quarks or even of nuclear physics, for that matter. The
reductionist's hope is that {\em the principles} governing all phenomena
stem from a fundamental underlying theory which, in turn,  would enable us to
derive seemingly fundamental concepts from a deeper origin. The
derivation of the empirical laws governing the behaviour of ideal gases
from kinetic theory provides a simple illustration of this. A different example
is the derivation of the Stefan-Boltzmnn law for the emissive power of a
blackbody (which also enables us to write Stefan's constant in terms of the
more fundamental Planck's constant). The derivation of the magnetic
susceptibility and polarizability of mono-valent gases in terms of
atomic properties of the corresponding atoms provides an illustration of
how ``long-distance characteristics'' -- in this example, some bulk properties of
gases -- can be obtained from the underlying microphysics. Continuing in
this vein, we may hope that in the future, some of the many disparate
parameters of the Standard Model which has been remarkably successful in
describing data up to distance scales down to (100~GeV)$^{-1}$, will be
derived from an underlying (more) fundamental theory that includes 
a detailed description of new degrees
of freedom with mass scales (much) higher than 100~GeV.

It is perhaps worth emphasizing that realizing such a top-down program
may prove very difficult, even in principle, because the low
energy theory may turn out to be sensitive to physics at all
energy scales. Although most of us implicitly assume that very high
energy scale degrees of freedom decouple from low energy physics, it
remains logically possible that this may not be the case. It could, for
instance, be that the multiplicity of massive states grows so rapidly
with mass, that even though the effect of any individual state is
negligible, their collective effect remains at low energy. In this case,
one would have to know the detailed physics at all energy scales to
realize the top-down program.

The other possibility is that low energy physics is insensitive to the
details of high scale physics because the effects of the latter are
suppressed by a power of the high scale $\Lambda$. This view provides a
rationale for the success that renormalizable relativisitic quantum
field theories have enjoyed in the describing strong and electro-weak
phenomena today, and makes a strong case that any mass scale associated
with unknown degrees of freedom lies well above the highest energies
accessible today, so that the effect of non-renormalizable operators is
sufficiently suppressed.

There is, however, an associated issue, brought to the forefront by the
discovery of the first (seemingly) elementary spin-zero particle at the
CERN collider with attributes remarkably consistent with those of the
Higgs boson of the Standard Model~\cite{atlas_h,cms_h}. 
Although, as we just said, low energy
phenomena are essentially independent of $\Lambda$, the {\it
dimensionful} parameters of the {\em renormalized} theory are generally
speaking sensitive to
the high scale $\Lambda$, and hence to the physics at high energy scales. For
instance, in a generic quantum field theory, the radiative corrections
to the squared mass of an elementary spin-zero particle take the form,
\be
m_{\phi}^2 - m_{\phi 0}^2 = C_1 \frac{g^2}{16\pi^2}\Lambda^2 + 
C_2 \frac{g^2}{16\pi^2}m_{\rm low}^2 \log\left(\frac{\Lambda^2}{m_{\rm low}^2}\right) 
+C_3 \frac{g^2}{16 \pi^2}m_{\rm low}^2\;.
\label{eq:generic}
\ee
The $C_3$  term could also include  ``small logarithms'' $\log(m_{\rm
low}^2/m_\phi^2)$ that we have not exhibited. 
We see the well-known quadratic sensitivity of scalar mass parameters to
the scale $\Lambda$ where new massive degrees of freedom that couple to
the SM reside; {\it e.g.} $\Lambda = M_{\rm GUT}$ when the SM is
embedded in a Grand Unified framework.\footnote{We stress that $\Lambda$
here is not {\it a regulator} associated with divergences that occur in
loop calculations in quantum field theory. Rather, it is the mass scale
associated with new particles with large couplings to the Higgs boson, a
point also made explicitly in Ref.~\cite{tait}. From this viewpoint, and
tempting though it is, we would not logically be able to associate
$\Lambda$ with $M_{\rm Planck}$, the scale at which the effects of
gravity become important. We do not really know quantum
gravitational dynamics and, in particular, do not know that there are
associated new particles with significant couplings to the Higgs boson.
See also Ref.~\cite{drees}.}  In Eq.~(\ref{eq:generic}), $m_{\phi}$ is
the physical mass of the quantum of the field $\phi$, $g$ is the typical
coupling of the field $\phi$, $m_{\phi 0}$ is the corresponding mass
parameter in the Lagrangian, $16\pi^2$ is a loop factor, and $C_i$ are
dimensionless coefficients that aside from spin, colour and other
multiplicity factors are numbers ${\cal O}(1)$.  Finally, $m_{\rm low}$
denotes the highest mass scale in the low energy theory, while $\Lambda$
is the scale at which this effective theory description becomes invalid
because the effects of heavy states not included in the Lagrangian that
provides a description of physics at low energies become important.  If
$\Lambda \gg m_{\rm low}$, unless $g$ is also tiny, the first term
dominates the corrections. Moreover, in order for the physical mass
$m_{\phi}$ to be at its fixed value in the low energy theory, it must be
that there are large cancellations between $m_{\phi 0}^2$ and the
$\Lambda^2$ term in Eq.~(\ref{eq:generic}). This quadratic sensitivity
of the radiative corrections to the squared mass parameter of elementary
spin-zero fields leads to the {\em fine-tuning problem} in the Standard
Model (SM) \cite{gild} when the SM is embedded into a Grand Unified
Theory. We stress that this is not a logical problem in the sense it
does not render the theory inconsistent, nor a practical problem that
precludes the possibility of making precise predictions using the SM. It
is only a problem in the sense that seemingly unrelated quantities in
Eq.~(\ref{eq:generic}) --- the mass parameter $m_{\phi 0}^2$ of the low
energy Lagrangian and contributions from radiative corrections from very
massive degrees of freedom governed by very different physics --- need a
cancellation of many orders of magnitude if $\Lambda \sim M_{\rm
GUT}$. Why should two quantities with very different physical origins
balance out with such exquisite precision?

The remarkable ultra-violet properties of softly broken supersymmetric
(SUSY) theories, with SUSY broken near the weak scale, ensure that the
low energy theory is at most logarithmically sensitive to high scale
(HS) physics, {\it i.e.}, that the $C_1$ term in Eq.~(\ref{eq:generic})
is absent. This led to the realization \cite{hier} that weak scale SUSY
potentially solves the {\it big gauge hierarchy problem} endemic to the
Standard Model (SM) \cite{gild} embedded into a GUT framework, and
provided much impetus for its study over the last three decades.  The
recent discovery of a Standard Model (SM)-like Higgs boson with mass
$m_h\simeq 125-126$~GeV \cite{atlas_h,cms_h} at the LHC seemingly
provides support for the simplest SUSY models of particle physics
\cite{wss,books} which had predicted $m_h\sim 115-135$~GeV
\cite{mhiggs}. However, no sign of supersymmetric matter has yet been
found at the LHC, resulting in mass limits $m_{\tg}\gtrsim 1.5$~TeV (for
$m_{\tg}\simeq m_{\tq}$) and $m_{\tg}\gtrsim 1$~TeV (for $m_{\tg}\ll
m_{\tq}$)\cite{atlas_susy,cms_susy}. Naively, this pushes up the SUSY
scale $m_{\rm low}$ to beyond the TeV range. If $\Lambda$ is not much
above the SUSY scale, the $C_{2,3}$ terms in Eq.~(\ref{eq:generic}) each
have a scale $\sim (100 \ {\rm GeV})^2$, which is comparable to the
observed value of the Higgs boson mass for $m_{\rm low} \lesssim
1-2$~TeV, and no large cancellations are necessary. However, one of the
most attractive features of supersymmetric theories is that they can be
perturbatively valid up to energy scales as high as $M_{\rm GUT}$ at
which the measured values of the three SM gauge couplings appear to
unify. In this case, $\Lambda \sim M_{\rm GUT} \simeq 2\times
10^{16}$~GeV, so that the $C_2$ term becomes two orders of magnitude
larger than (100~GeV)$^2$, requiring cancellations at the percent level
to obtain measured value of the Higgs boson mass. This need for
fine-tuning is what has been termed as the {\it Little Hierarchy
Problem}, to be contrasted with the {\it Big Hierarchy Problem} that is
solved by the introduction of weak scale SUSY as we mentioned earlier.

Fine-tuning in the Minimal Supersymmetric Standard Model (MSSM) is
seemingly exacerbated because experiments at the LHC have discovered a
SM-like Higgs boson with a mass at 125-126~GeV, well beyond its
tree-level upper bound $m_h \le M_Z$.  Radiative corrections can readily
accommodate this, but only with top squark masses beyond the TeV scale
along with large mixing\cite{h125}.  Since top squarks have large Yukawa
couplings to the Higgs boson, it has been argued that naturalness
considerations prefer $m_{\tst_{1,2}},m_{\tb_1}\lesssim
500$~GeV\cite{kn,ns,ah}. We will return to this issue below.

We recognize the inherent subjectivity of the notion of
naturalness. However, in order to decide whether one model is more
natural than another, we need to introduce a measure of fine-tuning.  As
we discuss in the next section, this is traditionally done by checking
the sensitivity of $M_Z^2$ rather than the Higgs mass as in
Eq.~(\ref{eq:generic}), to the model parameters. Since both gauge and
Higgs boson masses arise dynamically from the scalar potential, the
corresponding sensitivities are not unrelated.  

In the next section, we compare and contrast three different sensitivity
measures, $\delew^{-1}$, $\delhs^{-1}$ and $\delbg^{-1}$ that have been
the subject of discussion in the literature. While much of what we
say here and in the rest of this paper is a review, our perspective
differs from that of other authors. We emphasize that
while not a fine-tuning measure, $\delew$ (which is essentially
determined by the particle spectrum) is nonetheless a very
useful quantity because $\delew^{-1}$ quantifies the minimum fine-tuning
present in any theory with a specific spectrum. We also highlight
the importance of incorporating correlations between various
parameters in dicussions of fine-tuning, something  ignored in many
generic analyses. In Sec.~\ref{sec:example}, we
provide simple examples that suggest that if we can find HS theories
with specific correlations amongst parameters, the value of the
traditional fine-tuning measure $\delbg^{-1}$ would automatically be
close to $\delew^{-1}$. We then set up the
radiatively driven natural SUSY framework that we advocate 
for phenomenological analyses of natural models of SUSY, and review its
phenomenological implications in Sec.~\ref{sec:phen}. We conclude with
our perspective and a brief summary in Sec.~\ref{sec:concl}.

\section{Quantifying fine-tuning} 
\label{sec:measures}

The inherent subjectivity of the notion of fine-tuning is reflected in
the fact that there is no universally accepted criterion for when a theory
is fine-tuned. Everyone agrees that a model is natural if its
predictions can be obtained without the need for large cancellations
between various {\em independent} contributions that are combined to
obtain the predicted value of any quantity: see {\it e.g.} our
discussion following (\ref{eq:generic}). As we will see below, the
differences between various fine-tuning measures originate in whether
all truly independent contributions are really included, and (to a
lesser degree) on the sensitivity measure used. Our purpose is to
address whether supersymmetric models can be natural in light of what we
have learnt from LHC8 data.  Furthermore, we will limit our discussion
to the Minimal Supersymmetric Standard Model (MSSM) since part of our
goal is to examine whether naturalness considerations unequivocally
force us to consider extended frameworks. 

With this in mind, we discuss three different fine-tuning measures that
have received attention in the recent literature. As we have noted above,
the predicted value of $M_Z^2$ obtained from the minimization of the
one-loop-corrected Higgs boson potential
\be 
\frac{M_Z^2}{2} = \frac{m_{H_d}^2+\Sigma_d^d - (m_{H_u}^2+\Sigma_u^u) \tan^2\beta}{\tan^2\beta -1} -\mu^2,
\label{eq:mZsSig}
\ee 
is the starting point for most discussions of fine-tuning
\cite{ellis,bg,CCN}.  This expression is obtained using the weak scale
MSSM Higgs potential and all parameters in Eq.~(\ref{eq:mZsSig}) are
evaluated at the scale $Q=M_{SUSY}$.  The $\Sigma$s in
Eq.~(\ref{eq:mZsSig}), which
arise from one loop corrections to the Higgs potential, are the analogue
of the $C_3$ term in (\ref{eq:generic}). Explicit forms
for the $\Sigma_u^u$ and $\Sigma_d^d$ are given in the Appendix of
Ref.~\cite{rns}.

\subsection{$\Delta_{\rm EW}$}

Requiring that the observed value of $M_Z^2$ is
obtained without large cancellations means that each of the 
various terms on the right-hand-side of Eq.~(\ref{eq:mZsSig}) 
has to be comparable to $M_Z^2$ in magnitude. 
Thus the fine tuning in Eq.~(\ref{eq:mZsSig}) can be quantified
by $\delew^{-1}$, where \cite{ltr,sugra,rns}
\be 
\Delta_{\rm EW} \equiv max_i \left|C_i\right|/(M_Z^2/2)\;. 
\label{eq:delew}
\ee 
Here, $C_{H_d}=m_{H_d}^2/(\tan^2\beta -1)$,
$C_{H_u}=-m_{H_u}^2\tan^2\beta /(\tan^2\beta -1)$ and $C_\mu =-\mu^2$.
Also, $C_{\Sigma_u^u(k)} =-\Sigma_u^u(k)\tan^2\beta /(\tan^2\beta -1)$
and $C_{\Sigma_d^d(k)}=\Sigma_d^d(k)/(\tan^2\beta -1)$, where $k$ labels
the various loop contributions included in Eq.~(\ref{eq:mZsSig}).  We
immediately see that any upper bound on $\delew$ that we impose from
naturalness considerations necessarily implies a corresponding limit on
$\mu^2$. Thus higgsino masses are necessarily bounded from above in any
theory with small values of $\delew$.\footnote{This reasoning fails if
the dominant contribution to the higgsino mass arises from SUSY breaking
\cite{sundrum} and not from $\mu$. If there are no singlets that couple
to the higgsinos, such a contribution would be soft. However, in all HS models
that we are aware of, the higgsino masses have a supersymmetric
origin. }

Before proceeding further, we remark that $\delew$ as defined here
entails only weak scale parameters (see also Ref.~\cite{perel}) and so
has no information about the $\log\Lambda$ terms that cause weak scale
physics to exhibit logarithmic sensitivity to HS physics as discussed in
Sec.~\ref{sec:intro}.  For this reason $\delew$ is {\em not a
fine-tuning measure} in the underlying HS theory, as already noted in
Ref.~\cite{rns}. It is nonetheless very useful because, as noted below,
$\delew^{-1}$ yields a {\em lower bound} on the fine-tuning in any HS
theory with a given SUSY spectrum. Moreover, we will see in
Sec.~\ref{sec:example} that this bound can be saturated in an
appropriate HS theory with the same spectrum. For now we turn our
attention to $\delhs$ which includes the information of the large
logarithms in its definition.

\subsection{$\delhs$}

The large logarithms that we have been discussing remain hidden in
 Eq.~(\ref{eq:mZsSig}) because we have written this condition in terms
 of the parameters of the theory renormalized at the weak scale. To make
 these explicit, we rewrite the {\it weak scale} parameters
 $m_{H_{u,d}}^2$ and $\mu$ that appear in Eq.~(\ref{eq:mZsSig}) in terms of the
 parameters of the HS theory as,
\be m_{H_{u,d}}^2=
 m_{H_{u,d}}^2(\Lambda) +\delta m_{H_{u,d}}^2; \qquad
 \mu^2=\mu^2(\Lambda)+\delta\mu^2 \;, \ee 
where $m_{H_{u,d}}^2(\Lambda)$
 and $\mu^2(\Lambda)$ are the corresponding parameters renormalized at
 the high scale $\Lambda$.  In terms of the high scale parameters, the
 minimization condition now takes the form,
\be \frac{M_Z^2}{2} = \frac{(m_{H_d}^2(\Lambda)+ \delta m_{H_d}^2 +
\Sigma_d^d)-(m_{H_u}^2(\Lambda)+\delta
m_{H_u}^2+\Sigma_u^u)\tan^2\beta}{\tan^2\beta -1}
-(\mu^2(\Lambda)+\delta\mu^2)\;.
\label{eq:mZs_hs}
\ee 
The $\delta m_{H_{u,d}}^2$ and $\delta\mu^2$ terms contain the
$\log{\Lambda^2\over m_{\rm low}^2}$ factors that appears in the $C_2$ term in
(\ref{eq:generic}). Various authors \cite{kn,ah,ns} have argued that
this leads to rather stringent upper bounds on sparticle -- most notably top
squark -- masses from naturalness considerations. We will see below that
natural models of SUSY with top squarks beyond the reach of the LHC are
perfectly possible.

We can now define a fine-tuning measure that encodes the
information about the high scale origin of the parameters in a manner
analogous to the definition of $\delew$ above by now requiring \cite{sugra}
that none of the terms on the right-hand-side of Eq.~(\ref{eq:mZs_hs})
are much larger than $M_Z^2$. 
The high scale fine-tuning measure $\Delta_{\rm HS}$ is thus defined to be
\be 
\Delta_{\rm HS}\equiv max_i |B_i |/(M_Z^2/2)\;, 
\label{eq:hsft} 
\ee
with $B_{H_d}\equiv m_{H_d}^2(\Lambda)/(\tan^2\beta -1)$, 
$B_{\delta m_{H_d}^2} \equiv \delta m_{H_d}^2/(\tan^2\beta -1)$, {\it etc.}

In models such as mSUGRA~\cite{msugra}, whose domain of validity
extends to very high scales, because of the large logarithms one would
expect that the $B_{\delta m_{H_u}^2}$ contributions to $\Delta_{\rm HS}$
would typically  be much larger than any contributions to $\Delta_{\rm EW}$.
The reason is that the term $m_{H_u}^2$ evolves from large $m_0^2$ through zero
to negative values in order to radiatively break electroweak
symmetry. Put differently, the loop terms $\delta m_{H_{u,d}}^2$ in
Eq.~(\ref{eq:mZs_hs}) are typically much larger than the loop terms
in Eq.~(\ref{eq:mZsSig}) because of the presence of the large logarithm, and
we typically have,
\be
\delhs \gg \delew\;.
\label{eq:inequal1}
\ee 
Large cancellations between, for instance, $m_{H_u}^2(\Lambda)$
and $\delta m_{H_u}^2$ that result in small $\delew$ will nonetheless yield
a large value of $\delhs$.

Before closing this section, we note a potential pitfall of using
$\delhs$ as a measure of fine-tuning.  Although $\delhs$ is a sensible
measure of fine-tuning in a {\em generic} HS theory in that it captures
effects of the HS origin of the underlying model parameters, it does not
take into account the fact that in models with a small number of
parameters, the various terms on the right-hand-side of
(\ref{eq:mZs_hs}) could be correlated. In this case, there could be
automatic cancellations between the various terms that $\delhs$ does not
incorporate. In models where such cancellations occur
automatically,\footnote{Cancellations between $m_{H_u}^2(\Lambda)$ and
$\delta m_{H_u}^2$ are guaranteed for specially chosen values of
$m_{H_u}^2(\Lambda)$ in the NUHM2 model. The HB/FP region of mSUGRA
\cite{fp} and its generalizations \cite{sanford}, and the
mixed-modulus-anomaly-mediated SUSY breaking model (also referred to as
mirage mediation models) \cite{nilles} provide other examples of such
(partial) cancellations.}  using $\delhs$ could erroneously lead us to
infer that the model is fine-tuned. The possibility that correlations
among parameters can lead to reduced fine-tuning has been mentioned in
Ref.~\cite{reduce,CCN,antusch}. For another possibility, see 
Ref.~\cite{hardy}.

\subsection{$\Delta_{\rm BG}$}

The correlations that could be the pitfall of $\delhs$ as a fine-tuning
measure are most easily implemented in the traditional fine-tuning
measure $\Delta_{\rm BG}$ \cite{ellis,bg,dg}, defined as the
fractional change in the output value of $M_Z^2$ given by
(\ref{eq:mZsSig}) relative to the corresponding change in the input
parameters, 
\be 
\Delta_{\rm BG}= max_i|c_i|
\equiv max_i\left|\frac{a_i}{M_Z^2}\frac{\partial M_Z^2}{\partial
  a_i}\right|\;.
\label{eq:DBG}
\ee
Here, the $a_i$'s are the underlying parameters of the theory. These
would be the weak scale parameter set in the case of the pMSSM, in which
case $\delbg$ would be close to $\delew$, or the HS parameter set for
models such as mSUGRA. We would expect that in the latter case, aside
from the possibility of correlations discussed in the previous
paragraph, $\delbg$ and $\delhs$ will be very strongly
correlated.\footnote{$\delbg$ would equal to $\delhs$ if $M_Z^2$ depends
linearly on the model parameters $a_i$, and radiative corrections
embodied in $\Sigma$ are ignored.}  Remember, however, that for the
evaluation of $\delbg$, we need to combine terms on the right-hand-side
of Eq.~(\ref{eq:mZs_hs}) to calculate the sensitivity coefficients
$c_i$, above. If this combination results in cancellations because of
underlying correlations between HS parameters of the theory the
corresponding $c_i$, and concomitantly also $\delbg$, will automatically
reduce, whereas $\delhs$ (which does not know about these cancellations)
remains unchanged.  We thus generically expect that,\footnote{This
presumes that the dominant terms in $M_Z^2$ are (approximately) linear
in the parameters $a_i$. The semi-analytic formulae in
Eq.~(\ref{eq:mHu}) below show that this is indeed the case, except for
$a_i = m_{1/2}$. In the case that the sensitivity coefficient
$c(m_{1/2})$ is the largest, $\delhs$ can be twice as large as $\delbg$
because $M_Z^2$ is quadratic in $m_{1/2}$. }
\be 
\delew \le \delbg \lesssim \delhs.
\label{eq:inequal}
\ee

To make explicit what cancellations we are referring to, we see that for
moderate to large values of $\tan\beta$, we can write
Eq.~(\ref{eq:mZsSig}), to a good approximation, as 
\[
\frac{1}{2}M_Z^2
\simeq -m_{H_u}^2 -\mu^2.
\]
 The weak scale values of $m_{H_u}^2$ and
$\mu^2$ that appear above can be written in terms of the HS parameters
using the semi-analytic solutions to the one-loop renormalization group
equations \cite{munoz}.  For instance, for $\tan\beta =10$, we have
\cite{abe,martin,feng},
\bea 
-2\mu^2(m_{\rm weak}) &=& -2.18\mu^2\;, \\ 
-2m_{H_u}^2(m_{\rm weak})
  &=& 3.84 M_3^2+0.32M_3M_2+0.047M_1M_3-0.42 M_2^2 \nonumber 
 \\ &&+0.011 M_2M_1-0.012M_1^2-0.65M_3A_t-0.15 M_2A_t \nonumber 
 \\ &&-0.025M_1 A_t+0.22A_t^2+0.004M_3A_b\nonumber 
 \\ &&-1.27m_{H_u}^2 -0.053 m_{H_d}^2\nonumber 
 \\ &&+0.73m_{Q_3}^2+0.57 m_{U_3}^2+0.049m_{D_3}^2-0.052 m_{L_3}^2+0.053m_{E_3}^2 \label{eq:mHu}
 \\ &&+0.051m_{Q_2}^2-0.11 m_{U_2}^2+0.051m_{D_2}^2-0.052 m_{L_2}^2+0.053 m_{E_2}^2\nonumber 
 \\ &&+0.051m_{Q_1}^2-0.11 m_{U_1}^2+0.051m_{D_1}^2-0.052 m_{L_1}^2+0.053m_{E_1}^2 , \nonumber
\eea 
where the parameters on the right-hand-side are evaluated at the
GUT scale.\footnote{Eq.~(\ref{eq:mHu}) is written in the same convention as that used for input
parameters into ISAJET. The same convention is used throughout this paper. 
We warn the reader that in the convention of \cite{wss}, the signs of the 
$A_i M_j$ terms in (\ref{eq:mHu}) would have to be flipped.}
For other values of $\tan\beta$, the functional form on the
right-hand-side is the same except for somewhat different values of the
coefficients.  We can now use these to obtain the (approximate)
sensitivity coefficients and hence $\delbg$, using the semi-analytic
approximations in (\ref{eq:mHu}) above, assuming that the two-loop
effects and the radiative correction terms $\Sigma_u^u$ are small. We
will return to the validity of these approximations in Sec.~\ref{sec:example}.

The sensitivity coefficients depend on the underlying parameters of the
model. In the much-studied mSUGRA/CMSSM framework, the gaugino mass
parameters all unify to a common parameter $m_{1/2}$ at the high scale
(usually taken to be $M_{\rm GUT}$) while {\em all} scalar masses are
assume to unify to $m_0$, and the trilinear couplings to $A_0$. In this
case, because the HS model parameters are strongly correlated,
Eq.~(\ref{eq:mHu}) collapses to,
\[
-2m_{H_u}^2(m_{\rm weak}) = 3.78m_{1/2}^2-0.82 A_0 m_{1/2} + 0.22A_0^2
+0.013m_0^2   \qquad ({\rm mSUGRA}).
\]
We see that in the mSUGRA framework, the hallmark universality of scalar
mass parameters accidently leads to a tiny coefficient in front of
$m_0^2$. 
The smallness of this coefficient is an example of
cancellations between $m_{H_u}^2(\Lambda)$ and $\delta m_{H_u}^2$ that
occur, {\it e.g.} in the HB/FP region of mSUGRA {\it provided that
contributions from $m_{1/2}$ and $A_0$ terms as well as from the
radiative corrections $\Sigma_u^u$ are also small;} see, however,
Ref.~\cite{sugra}. In this parameter region, $\delbg$ is significantly
smaller than $\delhs$.

\begin{center}
\begin{figure}[tbh]{
\includegraphics[width=9cm,clip]{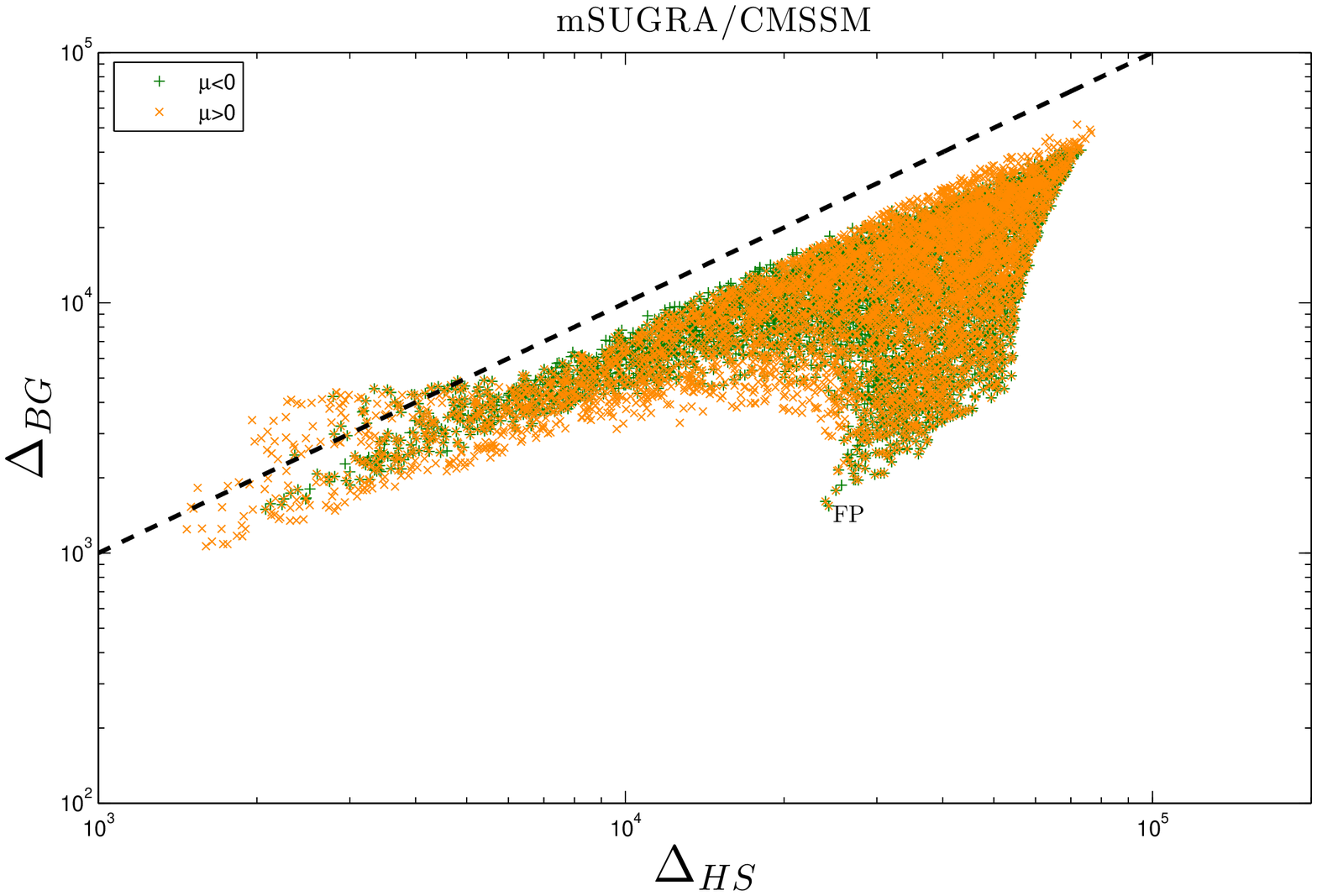}
\includegraphics[width=9cm,clip]{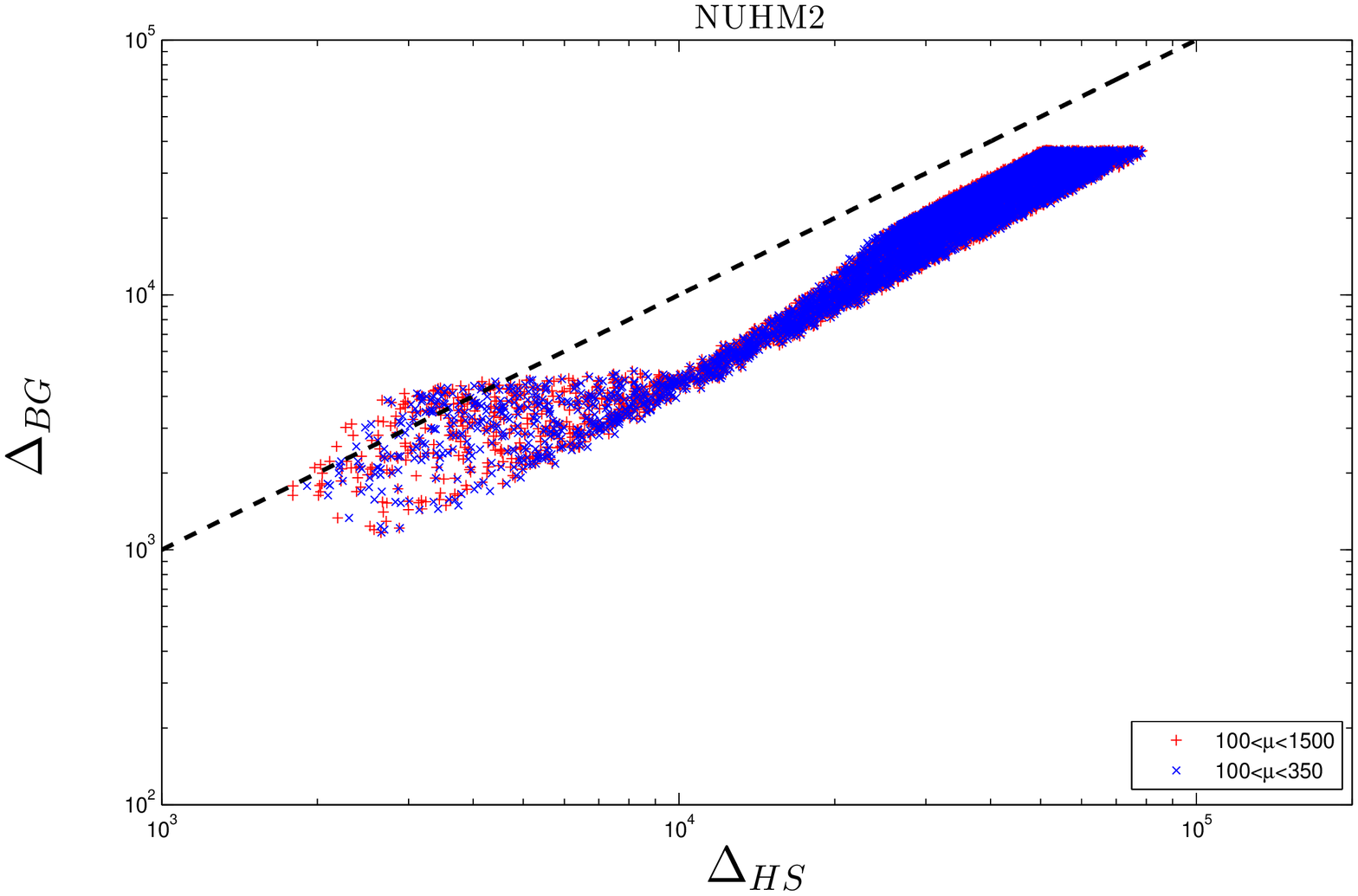}
\caption{Plot of $\delbg$ versus $\delhs$ from scans of ({\it a}) the
  mSUGRA parameter space (left frame), and ({\it b}) the NUHM2 model
  parameter space (right frame), as detailed in Ref.~\cite{meas}.}
\label{fig:bghs}}
\end{figure}
\end{center}
In Fig.~\ref{fig:bghs} we show $\delbg$ vs. $\delhs$ for a scan over the
parameter space of phenomenologically consistent points in ({\it a}) the
mSUGRA model, and ({\it b}) the parameter space of the non-universal
Higgs mass model (NUHM2) which is just mSUGRA except that the GUT-scale
Higgs mass parameters $m_{H_u}^2$ and $m_{H_d}^2$, or equivalently the
weak scale values of $\mu$
and $m_A$, are chosen to be independent of the mSUGRA parameters
\cite{nuhm2}. For details of the scan, we refer the reader to
Ref.~\cite{meas} from which this figure has been adapted.

We see that for both models, $\delhs$ and $\delbg$ are strongly
correlated as anticipated above, and that the inequality
(\ref{eq:inequal}) is satisfied. The handful of points where $\delhs <
\delbg <2\delhs $ are presumably for the cases where $c(m_{1/2})$ is the
largest of the sensitivity coefficients. While $\delbg$ is generally
comparable to $\delhs$, there is a subset of points in the mSUGRA case 
(marked FP) where $\delbg$ is substantially smaller than $\delhs$. This
is the hyperbolic branch/focus point region~\cite{fp} of mSUGRA where the
correlations between the parameters significantly reduce the fine-tuning
as discussed in the previous paragraph. We see nevertheless, that
$\delbg$ (as well as $\delhs$, of course) is always larger than $\sim
10^3$ so that both models would be considered fine-tuned to at least a
part-per-mille, if $\delbg^{-1}$ is used as the fine-tuning measure. 

We contrast this with $\delew$ which was shown \cite{sugra} to have a
minimal value of ${\cal O}(100)$ in the mSUGRA model (after the
imposition of LHC Higgs and sparticle mass constraints) but could be as
small as 10 in {\em special regions} of the NUHM2 parameter space
\cite{rns,meas}. Following our earlier discussion, we interpret this to
imply that {\em any theory} that leads to an mSUGRA-like sparticle
spectrum with only MSSM particles at the SUSY scale, will be fine-tuned
to at least the percent level, but leaves open the possibility of
finding a much less fine-tuned HS model that reduces to NUHM2 (with
specific correlations between NUHM2 parameters) as the effective theory
at a scale near $Q=M_{\rm GUT}$. We will address this in
Sec.~\ref{sec:example}.

\subsection{The utility of $\delew$}

We have discussed three quantities , $\delew$, $\delhs$ and $\delbg$
related to fine-tuning, that satisfy (\ref{eq:inequal}).  Of these,
$\delhs$ and $\delbg$ include information about potential enormous
cancellations that may be needed if there are states with mass scales
vastly greater than experimentally accessible energies that couple to SM
particles and their superpartners. In contrast, $\delew$ completely
disregards these cancellations since, by construction, it has no
sensitivity to ultra-violet physics. The inequality (\ref{eq:inequal})
shows that $\delew^{-1}$ provides a bound on the fine-tuning measure in
a generic quantum field theory in that measures the minimal fine-tuning
that is present for a given spectrum.\footnote{We are well aware that the
inequality $\delew \le \delbg$ need not hold in the strict mathematical
sense. An extreme, albeit contrived, example may be a
meta-theory where all mass parameters are determined by a single mass
scale $m$, so that $M_Z^2 = a m^2$, with $a$ fixed by the theory. In
this case $\delbg=1$, whereas $\delew$, as we have defined it, may well
be larger. (This is because our definition of $\delew$, like that of
$\delhs$, does not incorporate  correlations between parameters.)
In such a theory (if it exists) fine-tuning is a vacuous
concept. Despite this, we believe $\delew$ provides a useful bound on
the fine-tuning because it applies in all models where $M_Z^2$ receives
sizeable contributions from two or more {\em uncorrelated} terms enhanced by
$\log\Lambda$. This is indeed the case in many models.}  While a model
with a small value of $\delew$ is not necessarily free of fine-tuning, a
model with a large value of $\delew$ is always fine-tuned.  $\delhs$ and
$\delbg$ are usually comparable and differ from each other only when
there are correlations between HS parameters that lead to automatic
cancellations between terms on the right-hand-side of (\ref{eq:mZs_hs}),
or equivalently (\ref{eq:generic}). Inclusion of these correlations is
essential to obtain a true sense of fine-tuning in a particular model.

The utility of $\delew$ arises from the fact that it is essentially
determined by the weak scale spectrum \cite{rns}, {\it i.e.}, different
HS theories that lead to the same sparticle spectrum will yield nearly
the same value of $\delew$, even though these may have vastly different
values of $\delhs$ or $\delbg$. A small value of $\delew$ in, say, some
region of parameter space of the NUHM2 model, offers the possibility
that one may discover a HS theory with essentially the same spectrum
that simultaneously has a small value of $\delbg\sim\delew$. This HS
model (if it exists) will then be the underlying theory with low
fine-tuning. Since many broad features of the phenomenology are
determined by the spectrum, much of the phenomenology of the (unknown)
underlying theory is the same as those of the NUHM2 model with the same
spectrum.\footnote{Exceptions to this, would be phenomenological aspects
that are very sensitive to the mass correlations special to the NUHM2 model;
since these correlations would depend on the details of the model, the
NUHM2 model might not represent these faithfully. But many mass reaches
at the LHC, or SUSY contributions to the anomalous magnetic moment of
the muon or $b\to s\gamma$ (as long as there are no large cancellations
between various SUSY contributions), or even dark matter phenomenology
would be expected to be the same.} The underlying philosophy behind much
of our recent work~\cite{rns,rnslhc} is that the NUHM2 model acts as a
surrogate for the yet-to-be discovered theory with low fine-tuning.  The
other side of the same coin is that if we discovered superpartners and
found that these exhibited the spectrum of the mSUGRA model with
$m_h=125$-126~GeV, we would be forced to conclude that {\em any
underlying theory} that led to this spectrum would have to be 
fine-tuned~\cite{sugra}.

\section{How correlations (nearly) reduce $\delbg$ to $\delew$:\\ A simple example}
\label{sec:example}

We are led to conclude that $\delbg$ which includes information of both
UV physics and readily facilitates the inclusion of possible
correlations among HS parameters that lead to {\em automatic
cancellations} in (\ref{eq:mZs_hs}) is the optimal measure
of fine-tuning in quantum field theory. In contrast we have argued that
$\delew^{-1}$ yields a useful bound on the fine-tuning for a given
sparticle spectrum. In this section, we ask if the numerical value of
$\delbg$ would cause it to approach the value of $\delew$ once
correlations among the HS parameters are incorporated. For our study, we
adopt the NUHM2 model cases from Table~1 of Ref.~\cite{rns} that
resulted in low values of $\delew$.  Specifically, we have have:
\[
 {\bf Case A:}\ m_0=2.5~{\rm TeV}, m_{1/2}=0.4~{\rm TeV}; A_0=-4~{\rm TeV}; 
\tan\beta=10; m_A=1~{\rm TeV}; \mu=150~{\rm GeV},
\]
\[
{\bf Case B:}\ m_0=4~{\rm TeV}, m_{1/2}=1~{\rm TeV}; A_0=-6.4~{\rm TeV}; 
\tan\beta=15; m_A=2~{\rm TeV}; \mu=150~{\rm GeV}.
\]

Table~1 of Ref.~\cite{rns} shows that a change of $\sim$1\% in the GUT
scale values of $m_{H_u}^2$ caused $\delew$ to alter by $\sim 60-100$\%.
This had led us to suggest that if there was an underlying meta-theory
in which $m_{H_u}^2$ and $m_0^2$ were tightly correlated instead of
being independent parameters as in the NUHM2 model, this underlying
theory might not be fine-tuned.

In the NUHM2 model, Case A yields $\delbg=3168$ and $\delew=11.3$, while
for Case B we have $\delbg=8553$ and $\delew=16.9$. We immediately see
that since $\delew$ is two orders of magnitude smaller than $\delbg$, in
order to check whether the correlations indeed reduce $\delbg$ to (near)
$\delew$, the former would need to be computed to better than the
percent level. This precludes the use of semi-analytic 1-loop
expressions (\ref{eq:mHu}) that ignore two loop effects and also
evaluate the coefficients for a fixed value of $\tan\beta$ for the
computation of $\delbg$.\footnote{We emphasize that over much of the
parameter space of the NUHM2 model, the evaluation of $\delbg$ using
(\ref{eq:mHu}) (as was done in Ref.~\cite{meas}) will be reliable. Only
when the cancellation between the various terms approach the percent
level will this procedure become suspect.} We clearly need a different 
procedure to evaluate $\delbg$ in the meta-theory in which various NUHM2
parameters are correlated; {\it i.e.}, the meta-theory has fewer independent 
parameters than contained in the NUHM2 parameter set. 

We use the following multi-step procedure based on ISAJET\cite{isajet}
for a reliable  evaluation of $\delbg$:
\begin{enumerate}
\item Since the sensitivity coefficients needed for the evaluation of
$\delbg$ depend on GUT-scale parameters, for the NUHM2 point of interest
(for which ISAJET uses the weak scale values of $\mu$ and $m_A$ as
inputs), we first evaluate $m_{H_u}^2({\rm GUT})$ and $m_{H_d}^2({\rm
GUT})$ using two-loop renormalization group evolution instead of the
one-loop semi-analytic formulae mentioned above. 

\item We have created a program that uses these GUT-scale values of
  Higgs parameters together with other GUT-scale SUSY parameters to
  iteratively evaluate the SUSY spectrum. For this code, $|\mu|$ and
  $M_Z$ are outputs that (nearly) coincide with the input value of
  $|\mu|$ and the observed value of $M_Z$. We use the GUT-scale values
  of gauge and Yukawa couplings from the last iteration for this
  calculation.  The values of $\Sigma_u^u$ and $\Sigma_d^d$ are also
  re-evaluated.  

\item To evaluate the sensitivity coefficients that enter the computation 
  of $\delbg$, we now incrementally change each of the {\em independent}
  GUT-scale input parameters one-by-one (keeping all other parameters
  fixed) and reevaluate $M_Z^2$. The sensitivity coefficient is then
  obtained using $c_i = \frac{a_i}{M_Z^2}\frac{\delta M_Z^2}{\delta
  a_i}$. The largest of the sensitivity coefficients is taken as
  $\delbg$. Within the NUHM2 model, the parameters $m_0, m_{1/2}, A_0,
  m_{H_u}^2({\rm GUT})$ and $m_{H_d}^2({\rm GUT})$ are all independent,
  and so each one of these has a sensitivity coefficient that enters the
  evaluation of $\delbg$. As noted, this gives $\delbg =3168$ and
  $\delbg=8553$ for Cases A and B, respectively. The situation is quite
  different if the NUHM2 arises from a meta-theory in which the
  parameters are correlated as described below.

\item Next, motivated by our earlier studies, we imagine that the NUHM2
  is derived from a meta-model in which $A_0$ is not an independent
  parameter but is fixed in terms of $m_0$ by $A_0 = \xi_{A}m_0$, with
  $\xi_A \sim -1.6$. This correlation reduces $\tst_1$ contributions to
  $\Sigma_u^u$ and simultaneously raises $m_h$ to its observed
  value~\cite{ltr}.  In the meta-model, the sensitivity coefficient
  corresponding to $A_0$ should not be included during the evaluation of
  $\delbg$ because $A_0$ is not an independent parameter. For the two
  cases that we examined (and likely over much of parameter space), the
  value of $\delbg$ was fixed by sensitivity coefficients other than
  $c_{A_0}$, and so remains unchanged from its value in the NUHM2 model.

\item Recalling that the adjustment of the GUT-scale value of
  $m_{H_u}^2$ was key to obtaining a low value of $\delew$ in the NUHM2
  framework \cite{rns}, we assume that, like $A_0$, $m_{H_u}^2({\rm
  GUT})$ is also not an independent parameter in the meta-theory. Since
  the sensitivity to $m_{H_u}^2({\rm GUT})$ was dominant in the NUHM2
  model, viewing this as a dependent parameter can dramatically reduce
  $\delbg$.  Taking   $m_{H_u}^2({\rm GUT}) = \xi_H m_0^2$ with $\xi_H =
  1.64$ (1.70)  in Case A  (Case B) reduces $\delbg$ by about an order
  of magnitude.  

\item Finally, if we assume that the gaugino masses are also not
  independent but given by $m_{1/2}=\xi_{1/2}m_0$ with $\xi_{1/2} = 0.16$
  (0.25) in Case A (Case B), $\delbg$ drops by another order of
  magnitude. We emphasize that the spectrum and, in fact, all
  phenomenological predictions of this meta-theory will be identical to
  those of the NUHM2 model with the same parameters.

\end{enumerate}

The impact of these correlations between the parameters of the
meta-theory on $\delbg$ is illustrated in Table~\ref{tab:correl}.
\begin{table}
\begin{center}
\begin{tabular}{|l|r|r|}
\hline
Correlation & {\bf Case A}   &  {\bf Case B} \\
\hline
\hline
None    & 3168      &  8553 \\
$A_0=\xi_A m_0$, $m_{H_u}^2=\xi_H m_0^2$ & 257        & 1123 \\
$m_{1/2}=\xi_{1/2}m_0$  & 15.4 &  55 \\
$\delew$  & 11.3 &  $17$ \\
\hline
\end{tabular}
\caption{Values of $\delbg$ for the two cases of the NUHM2 model
introduced in the text. The first row shows the value of $\delbg$
without any correlations; in the second row we take $A_0$ and
$m_{H_u}^2({\rm GUT})$ to be determined by $m_0$ with $\xi_H=1.64$ for
Case A, and 1.70 for Case B, with $\xi_A=-1.6$ for both Cases. In the
third row we assume that the value of $m_{1/2}$ is also determined by
$m_0$ with $\xi_{1/2}=0.16$ (0.25) for Case A (Case B). The last row
shows the value of $\delew$.
\label{tab:correl}}
\end{center}
\end{table}
We see that in a meta-model with $\xi_A = -1.6$ which automatically
reduces the value of $\Sigma_u^u$, correlating the GUT-scale parameter
 $m_{H_u}^2$ reduces $\delbg$ by an order of magnitude. A reduction by
 another order of magnitude, leading to $\delbg$ not far from $\delew$,
 is obtained by  also correlating $m_{1/2}$.  The following remarks
 are worth noting.
\bi
\item It is clear that the small values of $\delbg$ in the penultimate 
row of Table~\ref{tab:correl} are the result of very substantial
cancellations between various contributions. This makes its evaluation
numerically delicate. Here, we have chosen the values of $\xi_H$ and
$\xi_{1/2}$ directly from Ref.~\cite{rns} without attempting to check
whether adjusting these will bring $\delbg$ yet closer to
$\delew$. Indeed, with our present code, we are unable to tell whether
the inequality $\delbg \ge \delew$ is saturated within numerical
error. The main new result is that, as anticipated, correlations among
high scale parameters substantially reduce $\delbg$, and if we are able
to find a meta-theory that results in these correlations, this theory
will have low fine-tuning.

\item The reader may be bothered by the fact that $\xi_H$ and
  $\xi_{1/2}$ change somewhat from Case A to Case B. However, this is
  not an issue since it is entirely possible that $A_0$, $m_{H_u}^2({\rm
  GUT})$ and $m_{1/2}$ are not correlated to just $m_0$ in the
  meta-theory; {\it i.e.}, the $\xi_\bullet$ could well be functions also
  of other
  parameters. The fact that the $\xi_\bullet$ are not widely different
  between cases, and are ${\cal O}(1)$ perhaps lends some support for
  our picture.
\ei

To recap, there are special regions of the parameter space of the NUHM2
model with small values of $\delew \sim 10-20$. In these regions, the
value of $\delhs$, or even $\delbg$ with all NUHM2 parameters treated
independently, is large $\sim 10^3$. However, if we assume that the
NUHM2 model is derived as the effective theory with (some of) its
correlated as described above, we find that the value of $\delbg$ drops
dramatically and assumes values not far from $\delew$. The parent
theory, if it exists, that gives rise to these correlations among NUHM2
model parameters will have much lower fine-tuning than in the NUHM2
model. We freely admit that we do not have any idea of how the required
correlations between parameters will arise -- this will surely require a
complete understanding of how supersymmetry is broken and how this
breaking is communicated to MSSM superpartners -- or even whether what
we are suggesting is possible. Our point is that we can consistently
speculate about such a possibility only in models where $\delew$ is
small. Since many aspects of the phenomenology are fixed only by the
super-partner spectrum, we can regard the NUHM2 framework with low value
of $\delew$ as a surrogate for the underlying (unknown) meta-theory with
low fine-tuning, and examine the experimental implications at the LHC
within this framework. This forms the subject of the next section.

\section{Radiatively Driven Natural Supersymmetry (RNS)}
\label{sec:rns}

It is clear from (\ref{eq:delew}) that a low value of $\mu^2/M_Z^2$ is a
necessary (though not sufficient) condition for obtaining a small value
of $\delew$. Since, aside from radiative corrections, (\ref{eq:mZsSig})
reduces to
\[
\frac{1}{2}M_Z^2 \simeq -m_{H_u}^2 -\mu^2,
\]
 for moderate to large
$\tan\beta$ it is clear that a weak scale value of $m_{H_u}^2$ close to
$M_Z^2$ guarantees a correspondingly small value of $\mu^2$.  This can
always be realized in the NUHM2 framework since $m_{H_u}^2({\rm GUT})$
is an adjustable parameter. From the perspective of the NUHM2 framework
this may necessitate a fine-tuning. However, as discussed a length in
the last section, it leaves open the possibility of finding a HS theory
with essentially the same mass spectrum that is fine-tuned at the level
of $\delew^{-1}$, not $\delbg^{-1}$ as computed in the NUHM2 model.

To find these low $\delew$ solutions, we perform scans of the NUHM2
parameter space as described in detail in Ref.~\cite{rns,rnslhc,meas}, requiring
that (1) electroweak symmetry is radiatively broken,  (2)~LEP2 and LHC
bounds on superpartner masses are respected, and (3)~that the value of
$m_h$ is consistent with the value of the Higgs boson
mass measured at the LHC. The low $\delew$ solutions of course have
low values of $|\mu|$, and generally have 
$A_0 \sim -(1-2)m_0$; this value typically leads to a cancellation of
the $\tst_1$  contribution to $\Sigma_u^u$ (the $\tst_2$ contribution is
suppressed if $m_{\tst_2} \sim (2.5-3)m_{\tst_1}$), and at the same time leads
to large intra-generational top squark  mixing that is required to raise
the Higgs mass to $\sim 125$~GeV. Since the required small value of
$|\mu|$ is obtained by $m_{H_u}^2$ being driven from its GUT scale
choice to close to $-M_Z^2$ at the weak scale, this scenario has been
referred to as {\em Radiatively Driven Natural Supersymmetry} (RNS).
It can be used a surrogate for an underlying natural model of
supersymmetry, and we urge its use for phenomenological analysis.

The RNS spectrum is characterized by:
\bi
\item the presence of four higgsino-like states $\tz_1,\tz_2$ and
  $\tw_1^\pm$ with masses in the  100-300~GeV range, and mass splitting
$\sim 10-30$~GeV between $\tz_2$ and the lightest supersymmetric
  particle  (LSP);

\item $m_{\tg} \sim 1.5-5$~TeV, with $\tz_{3,4}$ and $\tw_2^\pm$ masses
  fixed by (the assumed) gaugino mass unification condition;

\item $m_{\tst_1}=1-2$~TeV, $m_{\tst_2}, m_{\tb_{1,2}} \sim 2-4$~TeV; this
  is in contrast to many other studies that suggest that the stops should be
  in the few hundred GeV range, and so likely be accessible at the LHC. 

\item first and second generation sfermions in the 10~TeV range; this is
  not required to get low $\delew$, but compatible \cite{intra} with
  it. This choice ameliorates the SUSY flavour and CP problems
  \cite{decoup}, and also raises the proton lifetime \cite{proton}.
\ei

\section{Phenomenology}
\label{sec:phen}

We have seen that 100-300~GeV  charged and neutral higgsinos, with a mass gap
of 10-30~GeV with the LSP, are the hallmark of scenarios with $\delew
\lesssim 30$.  In this section, we present an overview of how SUSY
signals may be detected in such scenarios, highlighting those signatures
that may point to the underlying low value of $|\mu|$.

\subsection{LHC} 
\label{subsec:lhc}

\begin{figure}[tbh]{\begin{center}
\includegraphics[width=12cm,clip]{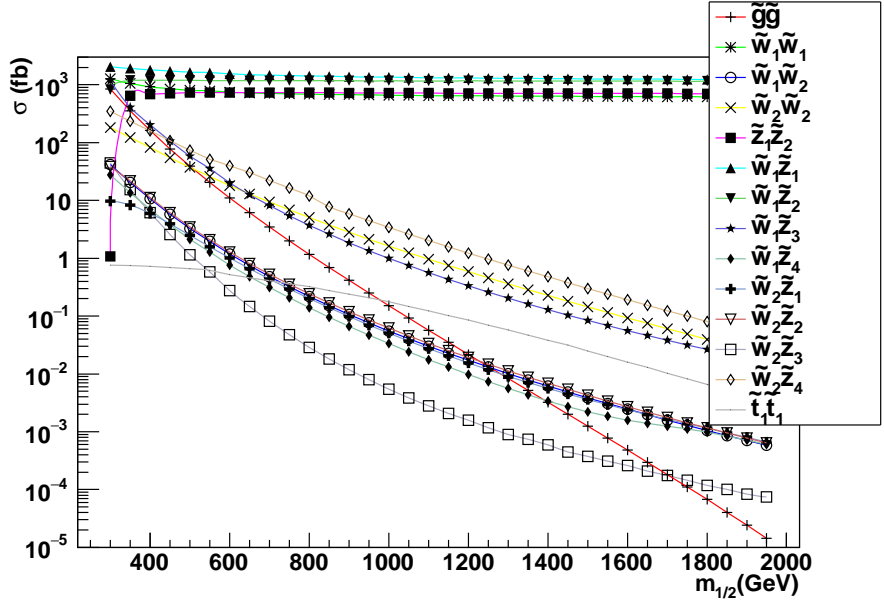}
\caption{Plot of various NLO sparticle pair production cross sections 
versus $m_{1/2}$ along the RNS model line (\ref{eq:mline}) for $pp$ collisions at 
 $\sqrt{s}=14$~TeV.
}
\label{fig:csec}\end{center}}
\end{figure}

Within the RNS framework, light higgsinos are likely to be the most
copiously produced superpartners at the LHC \cite{rnslhc}. This is
illustrated in Fig.~\ref{fig:csec} where we show various -ino production
cross sections (squarks and sleptons are assumed to be heavy as we
adopt the decoupling solution to the SUSY flavour problem) at LHC14.
The small energy release in their decay
makes their signals difficult to detect over SM backgrounds and we are
led to investigate other channels for discovery of SUSY.

{\em Gluinos:} Gluino pair production leads to the usual cascade decay
signatures in the well-studied multi-jet + multilepton channels. The
fact that lighter charginos and neutralinos are higgsino-like rather
than gaugino-like would affect the relative rates for topologies
with specific lepton multiplicity, but are unlikely to significantly
alter the reach which is mostly determined by the gluino production 
cross-section (which is essentially determined by the gluino and
first-generation squark masses). A study of the gluino reach within the
RNS framework shows that experiments at LHC14 should be sensitive to
$m_{\tg}$ values up to 1700~GeV (1900~GeV), assuming an integrated
luminosity of 300 (1000)~fb$^{-1}$. It may also be possible to extract
the value of $m_{\tz_2}-m_{\tz_1}$ from the end-point of the mass
distribution of opposite sign/same flavour dileptons from the
leptonic decays of $\tz_2$ produced in gluino decay cascades, if the mass
$\tz_2-\tz_1$ mass gap is large enough \cite{rnslhc}. We note,
however, that experiments at the LHC can discover gluinos only over part
of the range allowed by naturalness considerations. 

{\em Same Sign Dibosons:} If $m_{1/2}$ happens to be small enough so that
the bino and wino mass parameters are not hierarchically larger than
$|\mu|$, the two charginos and all four neutralinos will be mixed
gaugino-higgsino states with substantial mass gaps between the
heavier-inos and the LSP. Moreover, these states will all be
kinematically accessible at the LHC via electroweak production
processes, and we will be awash in multilepton signals with hadronic
activity only from QCD radiation. In this fortituous circumstance,
the gluino signal discussed above will also likely be detactable. 

The more typical scenario is when $|\mu| \ll M_{1,2}$ so that $\tw_1$
and $\tz_2$ are higgsino-like and only 10-30~GeV heavier than $\tz_1$,
$\tz_3$ is dominantly a bino, and $\tw_2$ and $\tz_4$ are winos. Because
squarks are heavy, and the bino does not have couplings to $W$ and $Z$
bosons, electroweak production of $\tz_3$ is dynamically
suppressed. However, winos have large ``iso-vector'' couplings to the
vector bosons so that wino cross sections can be substantial. Indeed we
see from Fig.~\ref{fig:csec} that $\tw_2^\pm\tw_2^\mp$ and $\tw_2\tz_4$
cross sections remain substantial for high values of
$m_{1/2}$.\footnote{The $\tw_1\tz_3$ cross section is also significant,
but falls more steeply with $m_{1/2}$ because the gaugino-higgsino
mixing becomes increasingly suppressed.} The large wino production
cross-section leads to a novel signature involving same-sign dibosons 
\begin{figure}[tbh]{\begin{center}
\includegraphics[width=10cm,clip]{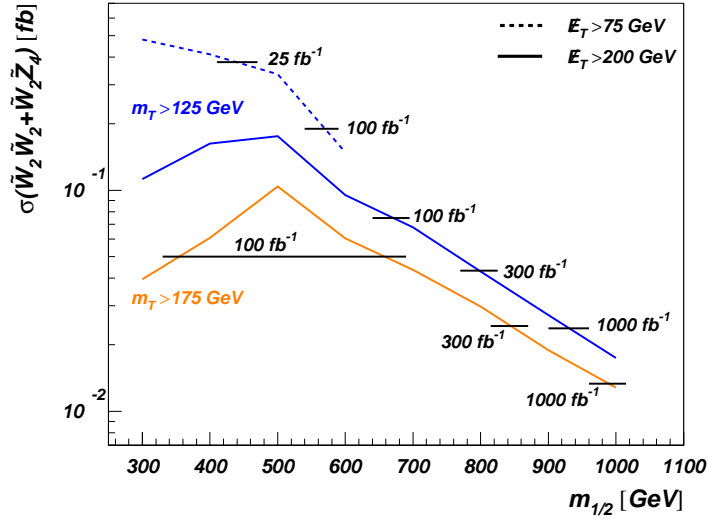}
\caption{Same-sign dilepton cross sections (in {\it fb}) at LHC14 after cuts vs. $m_{1/2}$
along the RNS model line (\ref{eq:mline}) from $\tw_2^\pm\tz_4$ and
$\tw_2^\pm\tw_2^\mp$ production and
calculated reach for 100, 300 and 1000~fb$^{-1}$.  The upper solid and
dashed (blue) curves requires $m_T({\rm min}) >125$~GeV while the lower solid
(orange) curve requires $m_T({\rm min}) >175$~GeV. The signal is observable
above the horizontal lines.
}
\label{fig:ssdbreach}\end{center}}
\end{figure}
produced via the processes $pp \to
\tw_2^{\pm} (\to W^{\pm}\tz_{1,2})+\tz_4 (\to W^\pm\tw_1^\mp)$. The
decay products of the lighter chargino/neutralinos tend to be soft, so
that the signal of interest is a pair of same sign high $p_T$ leptons
from the decays of the $W$-bosons, with limited jet activity in the
event. This latter feature serves to distinguish this source from same
sign dilepton events that might arise at the LHC from gluino pair
production.  We mention that $pp \to \tw_2^\pm \tw_2^\mp$ production
(where one chargino decays to $W$ and the other to a $Z$) also makes a
non-negligible contribution to the $\ell^\pm\ell^\pm +\eslt$ channel
when the third lepton fails to be detected. We emphasize here that this
signal is a hallmark of all low $\mu$ models, if wino pair production
occurs at substantial rates at the LHC.

We refer the reader interested in the details of the analysis required
to separate the signal from SM backgrounds to Sec.~5 of
Ref.~\cite{rnslhc}. We only mention that a hard $\eslt$ cut and, very
importantly, a cut on 
\[
m_T^{\rm min} \equiv {\rm
  min}\left[m_T(\ell_1,\eslt), m_T(\ell_2,\eslt)\right]
\] 
are very
effective for suppressing the backgrounds relative to the signal. 
The $5\sigma$  reach of the LHC for an NUHM2 model line with,
\be 
m_0=5~{\rm TeV}, A_0 = -1.6m_0, \tan\beta=15, \mu=150~{\rm GeV}, m_A=
1~{\rm TeV}, 
\label{eq:mline}
\ee
chosen to lead to low $\delew$, is illustrated in
Fig.~\ref{fig:ssdbreach} as a function of the gaugino mass parameter
$m_{1/2}$.
We show results for relatively soft cuts (dashed lines) and
hard cuts on $\eslt$ and $m_T^{\rm min}$. We see that with 300~fb$^{-1}$
of integrated luminosity, experiments at the LHC will probe $m_{1/2}$
values up to 840~GeV, well in excess of what can be probed via cascade
decays of gluinos.

{\em Hard Trilepton Signals:} Since low $|\mu|$ models yield such a large
reach for winos, it is natural to ask how far the wino reach extends in
the canonical trilepton channel, {\it i.e.}, from the reaction $pp \to
\tw_2 + \tz_4 X \to W+Z +\eslt+ X$, long considered to be the golden
mode for SUSY searches~\cite{trilep}. Here the $\eslt$ arises from the
$\tw_1/\tz_{1,2}$ (whose visible decay products are very soft) daughters
of the winos. A detailed analysis \cite{rnslhc} shows that the reach via
this channel extends to $m_{1/2} = 500$ (630)~GeV for an integrated
luminosity of 300 (1000)~fb$^{-1}$, considerably lower than via the SSdB
channel.

{\em Four Lepton Signals:} Low $|\mu|$ models, however, offer the
possibility of $ZZ+\eslt$ events from $\tw_2^+\tw_2^-$ or $\tw_2^\pm \tz_4$
production, when both winos decay to $Z$ plus a light
chargino/neutralino. This leads to the possibility of a 4 lepton signal
at LHC14. The reach in this channel was also mapped out in
Ref.~\cite{rnslhc}, by requiring 4 isolated leptons with $p_T(\ell) >
10$~GeV, a $b$-jet veto (to reduce backgrounds from top quarks), and
$\eslt > \eslt({\rm cut})$. The value of $\eslt({\rm cut})$ was chosen
so as to optimize the signal relative to SM backgrounds from $ZZ,
t\bar{t}Z, ZWW, ZZW, ZZZ$ and $Zh(\to WW^*)$ production. Since the
background also includes a $Z$ boson, and also because one of the four
leptons in the signal occasionally arises as a leptonic daughter of the
lighter $\tw_1$ or $\tz_2$, requiring a lepton pair to reconstruct
$M_Z$, in fact, reduces the signal significance. It was found that in
low $|\mu|$ models, the LHC14 reach via the $4\ell$ search extends
somewhat beyond that in the trilepton channel. Indeed a signal in this
channel together with the SSdB signal could point to a SUSY scenario
with small value of $|\mu|$ and a comparatively larger wino mass, as
might be expected in RNS.

{\em Soft Trileptons:} The reader will remember from Fig.~\ref{fig:csec}
that higgsino pair production is the dominant sparticle production
mechanism at the LHC. This naturally leads to the question whether the
$e\mu\mu$ signal from $\tw_1\tz_2$ might be observable, since the CMS
and ATLAS experiments may be able to detect muons with $p_T(\mu)$ as
small as 5~GeV. With this in mind, we examined the shape of the mass
distribution of dimuons in the reaction $pp \to \tw_1(\to
e\nu\tz_1)+\tz_2(\to \mu^+\mu^-\tz_1)$ in Ref.~\cite{rnslhc}, with cuts
chosen to enhance the soft trilepton signal over large SM
backgrounds. The signal dimuons would all have a mass smaller than the
kinematic end point at $m_{\tz_2}-m_{\tz_1}$, while the background
distribution would be expected to be much broader. Indeed it was found
that there should be an enhancement of this distribution at small values
of $m(\mu^+\mu^-)$, so that a {\em shape analysis} may well reveal the
signal if $m_{1/2} < 400-500$~GeV, for $\mu=150$~GeV. For larger values
of $\mu/m_{1/2}$ the mass gap is so small that the resulting spectral
distortion is  confined to just one or two low mass bins. We conclude
that while the soft-trilepton signal is unlikely to be a discovery
channel, it could serve to strikingly confirm a SUSY signal in the SSdB
or multilepton channels, and most importantly, point to a small value of
$|\mu|$ if the parameters are in a fortituous mass range.

{\em Mono-jet and Mono-photon Signals:} Many authors have suggested that
experiments at LHC14 may be able to identify the pair production of LSPs
via high $E_T$ mono-jet or mono-photon plus $\eslt$ events, where the
jet/photon results from QCD/QED radiation. Many of these studies have
been performed using non-renormalizable contact operators for LSP
production. This overestimates the rates for mono-jet/mono-photon
production at high $E_T$ especially in models such as RNS where
$s$-channel $Z$ exchange dominates LSP pair production \cite{buch}. A
careful study of this signal for the case of the higgsino LSP,
incorporating the correct matrix elements as given within the RNS
framework, shows that the signal will be very difficult to extract above
the SM backgrounds, unless these can be controlled at the better than
the percent level \cite{mono}. This is largely because the jet/photon
$E_T$ distribution as well as the $\eslt$ distribution has essentially
the same shape for the signal and the background. Alternatively,
detection might be possible if the soft daughter leptons from the decays
of the higgsino-like $\tw_1$ and $\tz_2$ can serve to reduce the
background in events triggered by the hard jet and/or
$\eslt$.\footnote{This has been examined in Ref.~\cite{kribs} where the
authors suggest this is feasible, at least for a sizeable mass
gap. There were no explicit studies for a mass gap down to $\sim 10$~GeV
that would be possible in the RNS scenario.}

Table~\ref{tab:reach} summarizes the projected reach of LHC14 in terms 
of the gluino mass within the RNS framework that we advocate be used for
phenomenological analyses of natural SUSY.
\begin{table}
\begin{center}
\begin{tabular}{|l|r|r|r|r|}
\hline
 Int. lum. (fb$^{-1}$) & $\tg\tg$ &  SSdB & $WZ\to 3\ell$ &$4\ell$ \\
\hline
\hline
10   & 1.4 &   --  & -- & --\\
100  & 1.6 &  1.6 & -- & $\sim 1.2$\\
300  & 1.7 &  2.1 & 1.4& $\gtrsim 1.4$ \\
1000 & 1.9 &  2.4 & 1.6& $\gtrsim 1.6$ \\
\hline
\end{tabular}
\caption{Reach of LHC14 for SUSY in terms of gluino mass, $m_{\tg}$
  (TeV), assuming various integrated luminosity values along an RNS
  model line introduced in (\ref{eq:mline}).
\label{tab:reach}}
\end{center}
\end{table}
We see that for an integrated luminosity in excess of $\sim
100$~fb$^{-1}$ the greatest reach will be obtained via the SSdB
channel if we assume gaugino mass unification. More importantly, the
SSdB channel provides a novel way to search for a SUSY signal in {\em
  any natural model of supersymmetry} since, as we have emphasized,
the $\mu$ parameter needs to be small. In this case, there may be
striking confirmatory signals in the $4\ell$ and soft-trilepton
channels in addition to the much-discussed clean trilepton
signal from wino pair production.

\subsection{ILC} 

Because light higgsinos are $SU(2)$ doublets, they necessarily have
sizeable couplings to the $Z$ boson, and so should be copiously produced
in $e^+e^-$ colliders, unless their production is kinematically
suppressed. Since small $|\mu|$ is necessary for naturalness,
electron-positron linear colliders that are being envisioned for
construction are the obvious facility for definitive searches for
natural SUSY.  The issue, of course, is whether in light of the small
visible energy release in higgsino decays it is possible to pull out the
higgsino signal above SM backgrounds.
 
Here, we report preliminary results from an on-going study \cite{ilc}
of higgsino signals at an electron-positron linear collider with a
centre-of-mass energy of 250~GeV (ILC250) that is seriously being
considered for construction in Japan. For this study, we have chosen
the NUHM2 point with $m_0=7025$~GeV, $m_{1/2}=568$~GeV, $\mu=115$~GeV
with $\tan\beta=10$. This case has gluinos and squarks beyond the
current LHC reach (though it should be possible to find gluino and
even wino signals at LHC14), and has $m_{\tw_1}=117.3$~GeV,
$m_{\tz_2}=124$~GeV and $m_{\tz_1}=102.7$~GeV, with $\delew=14$. We
view this point (ILC1) as an ``easy case study'' because of the rather
large mass gap.

Backgrounds from $2\to 2$ production processes typically have visible
energies
near 250~GeV, except when neutrino daughters from the decay of produced
parents take away a large energy. In contrast, the signal has a visible
energy smaller than 50~GeV. Except for a small contribution from the
tail of the $e^+e^-\to WW$ production, the $2\to 2$ backgounds are
efficiently removed by a cut on $E_{\rm vis}$. Much more relevant are
backgrounds from ``two-photon'' processes, $e^+e^-\to e^+e^- f\bar{f}$
where the final state electrons and positrons carry off the bulk of the
energy and are lost down the beam-pipe.  However, except when
$f=c,b,\tau$ these events are back-to-back in the transverse plane and
have very low $\eslt$. After the additional cut $\eslt > 20$~GeV, the
signal from $e^+e^-\to \tw_1^+\tw_1^- \to q\bar{q}\tz_1+\ell\nu\tz_1$
production is readily visible in the $2j+1\ell$ channel with an
integrated luminosity of 100~fb$^{-1}$, where jets and
leptons are defined to have transverse energies bigger than 5~GeV. Beam
polarization is not necessary for this.

The signal from neutralino production\footnote{Pair production of
  identical higgsinos, $\tz_1\tz_1$ or $\tz_2\tz_2$ has a much smaller
  cross section as the coupling of the higgsinos to $Z$ is dynamically
  suppressed \cite{wss}.} via $e^+e^- \to \tz_1\tz_2 \to
\tz_1\ell^+\ell^-\tz_1$ is also detectable with additional cuts $\eslt> 15$~GeV,
$\Delta\phi(\ell\ell) < \pi/2$, as 
described in Ref.~\cite{ilc}. For this study, 90\% electron beam
polarization is required. Notice that despite the small
leptonic branching ratio for $\tz_2$ decay, the signal is best seen via the
leptonic decay of $\tz_2$. This is  because hadronic decays of
$\tz_2$ lead mostly to single jet event topologies.

Ref.~\cite{ilc} also examines a more challenging case, for a point
along the model-line (\ref{eq:mline})  with $m_{1/2}=1.2$~TeV. This
yields $m_{\tw_1} \simeq m_{\tz_2}=158$~GeV, and a mass gap with the
neutralino of just
$\sim 10$~GeV. This point is chosen because it has $\delew =28.5$, close to
what we consider the maximum for naturalness, and  a mass gap that
is near the minimum, consistent with naturalness considerations. For
this case, gluinos and all squarks (and likely also winos) are beyond
the LHC14 reach. 

For the heavier $\tw_1$ and $\tz_2$ mass for this case, we have
performed a study taking $\sqrt{s}=340$~GeV, just below the $t\bar{t}$
threshold. If the ILC is constructed, and its energy upgraded to study
the top quark threshold, we expect that there will surely be an ILC run
close to this energy.  The smaller mass gap leads to events with even
less visible energy than in the ILC1 case study just discussed. In this
case, requiring $E_{\rm vis} < 30$~GeV along with cuts on $\eslt$
and various jet and lepton angles in the
transverse plane suffices  to make the background
negligible, and render the signal observable at the $5\sigma$ level
\cite{ilc}. Indeed, since there may well be no visible signal at LHC14
in this difficult scenario, the ILC could well be a discovery machine
for SUSY! 

Although we have not performed a parameter space scan, the
fact that the signal can be extracted even in this nearly maximally
difficult RNS case strongly suggests that higgsino signal will be
observable at an $e^+e^-$ collider provided of course that the higgsinos
are kinematically accessible and that electron beam polarization is
available (for the neutralino signal). In fact, we are curently
investigating the prospects for mass measurements.

\subsection{Dark Matter} 

Since the LSP is likely higgsino-like in {\em all} models with natural
supersymmetry, it will annihilate rapidly (via its large coupling to the
$Z$ boson, and also via $t$-channel higgsino exchange processes) in the
early universe. As a result, in natural supersymmetry the measured cold
dark matter density cannot arise solely from {\em thermally produced
higgsinos} (remember that these are lighter than $\sim 300$~GeV) in
standard Big Bang cosmology. Dark matter is thus likely to be
multi-component. What is very interesting, however, is that because
naturalness considerations also impose and upper bound on $m_{\tg}$ and
corresponding limits on electroweak gaugino masses (via gaugino mass
unification), {\em the thermal higgsino relic density cannot be
arbitrarily small.}  Indeed, within the RNS framework, $\Omega_{\tz_1}
h^2$ must be between $\sim 0.004-0.03$, as shown by Baer, Barger and
Mickleson \cite{bbm_dm}. This has important implications for DM
detection experiments. Specifically, ton-size direct detection
experiments such as Xe-1Ton that probe the spin-independent nucleon LSP
cross section at the $10^{-47}-10^{-46}$~pb level will be sensitive to
entire range of the expected higgsino fraction. Thus, the outcome of
these experiments will have important ramifications for
naturalness.\footnote{We should remind the reader that there are the
usual caveats to this conclusion. For instance if physics in the sector
that makes up the remainder of the dark matter entails late decays that
produce SM particles, the neutralino relic density today could be
further diluted; see {\it e.g.} Ref.~\cite{howie_axion}.}

\section{Concluding Remarks} 
\label{sec:concl}

Naturalness is a measure of how sensitive low energy masses and
couplings are to the dynamics at hierarchically separated energy
scales, and so is an attribute of the underlying high scale
theory. The dynamics of the SM shows us that the Higgs boson mass
exhibits quadratic senstive to masses of new, heavy particles, if
these couple to the Higgs boson. This sensitivity is correspondingly
reduced if these particles have very weak couplings to the Higgs
sector, or couple only indirectly at the multi-loop level. In theories
that incorporate weak scale supersymmetry, the quadratic sensitivity
to the masses of particles at very high scales is reduced to
logarithmic sensitivity. In all these considerations, we agree with
most discussions of naturalness and fine-tuning in much of the
literature.  

Where we evidently differ from many authors is that we allow for the
possibility that model parameters that appear independent from our low
energy perspective may really correlated within the as yet undiscovered
underlying theory. These correlations, as we have argued in
Sec.~\ref{sec:example}, can easily change the fine-tuning measure by a
couple of orders of magnitude: our toy illustrations show that a theory
that appears to be fine-tuned at parts per ten thousand may actually be
fine-tuned at the few percent level.\footnote{The reader may object that
if we allow the possibility of correlations, one may even argue that the
Higgs mass parameter may not be fine-tuned in even the SM. While this is
logically possible, we are not imaginative enough to see how a quadratic
sensitivity to say the GUT scale would be reduced by many orders of
magnitude to a sensitivity at the percent or parts per mille level by
parameter correlations. Of course, a symmetry ({\it e.g.}  SUSY) does
just this, but more typically, symmetries are not preserved to yield
cancellations with the required precision.}  Ignoring these parameter
correlations is what leads to stringent limits on top squarks that are
usually advertized as the hallmark of natural supersymmetry
\cite{kn,ns,ah}.  Indeed, the measures $\delhs$ and $\delbg$ defined in
Sec.~\ref{sec:measures} both incorporate the sensitivity of $M_Z^2$ to
the physics of new particles at the high scale.  However, the effect of
parameter correlations is most simply encoded into $\delbg$, but is
technically difficult to incorporate into $\delhs$, because the
coefficients $B_i$ in (\ref{eq:hsft}) cannot easily be written in terms of the
model parameters in a simple way.

Whether or not a theory  is (or is not) natural clearly depends on
how very heavy particles couple to weak scale particles.  This is a
question of dynamics, and so cannot be answered by just looking
at the weak scale spectrum of the theory. For this reason, we cannot
regard $\delew^{-1}$ (which is essentially fixed by the spectrum) introduced
in (\ref{eq:delew}) as a measure of fine-tuning in the theory, 
in sharp contrast to the considerations in Ref.~\cite{meas}. 
Despite this, we agree with both Ref.~\cite{meas} and
\cite{tait} that fine-tuning considerations using the weak scale theory 
is very useful, albeit for different reasons from these authors.
We find that $\delew$ is extremely useful because
it serves as a bound on $\delbg$, the true fine-tuning measure: see
Eq.~(\ref{eq:inequal}). Any model that leads to a large value of
$\delew$ is certainly fine-tuned. A small value of $\delew$ in some
region of model parameter space does not guarantee the model is not
fine-tuned.  However, it leaves open the possibility that parameter
correlations required to zero in on this special part of parameter
space will, one day, be obtained from a
more fundamental underlying framework. Evaluation of $\delbg$ with
these parameter correlations incorporated, will then yield a value (close to)
$\delew$. However, until such time that we have such a theory, it is
useful to examine the low $\delew$ regions of the parameter
space of phenomenologically promising models because these serve as
surrogates for an  underlying theory with low fine-tuning, as
explained at the end of Sec.~\ref{sec:example}.  

\begin{figure}[tbh]{\begin{center}
\includegraphics[width=10cm,clip]{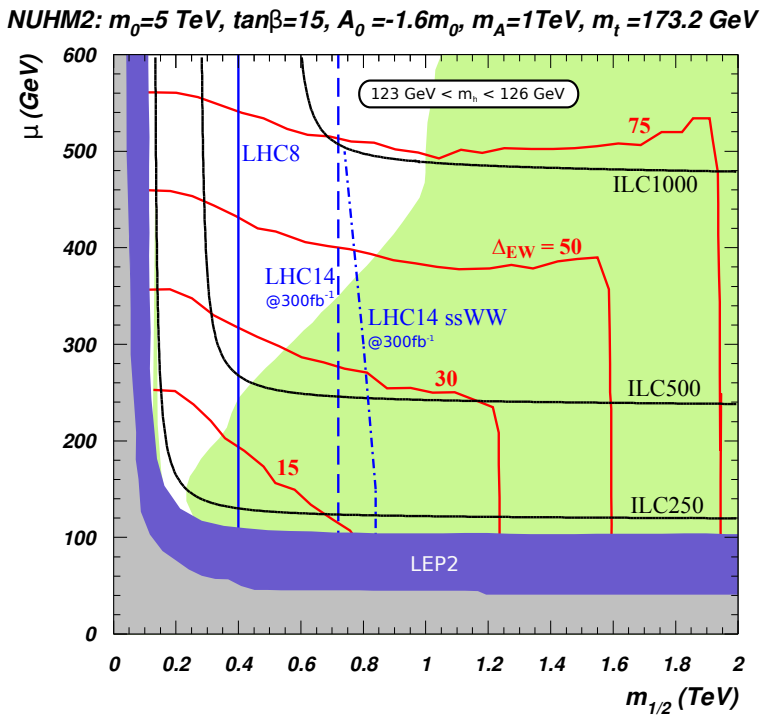}
\caption{Plot of $\delew$ contours (red) labelled by the value of
$\delew=15,30,50$ and 75 in the $m_{1/2}\ vs.\ \mu$
plane of NUHM2 model for $A_0=-1.6 m_0$ and $m_0=5$~TeV and $\tan\beta
=15$.  We show the region accessible to LHC8 gluino pair searches
(solid blue contour), and the region accessible to LHC14 searches with
300~fb$^{-1}$ of integrated luminosity (dashed and dot-dashed contours).
We also show the reach of various ILC machines for higgsino pair
production (black contours).  The very light-shaded (green) region has
$\Omega_{\tz_1}^{std}h^2<0.12$.  The dark  (light) shaded region along
the axes is
excluded by LEP2 (LEP1) searches for chargino pair production.  To aid
the reader, we note that $m_{\tg}\simeq 2.5 m_{1/2}$.}
\label{fig:pan}\end{center}}
\end{figure}

The RNS framework which, by construction, has a low value of $\delew$, provides
an explicit
realization of such a program. Since many phenomenological results are
sensitive to just the spectrum, these can be abstracted from the RNS
model. RNS phenomenology is discussed in Sec.~\ref{sec:phen}. In
Fig.~\ref{fig:pan}, we show the $m_{1/2}-\mu$ plane of the NUHM2 model
with large $m_0$ together with contours of $\delew$.
Above and to the right of the $\delew=30$ contour, we regard the
spectrum to be fine-tuned since the fine-tuning {\em must be} worse than
$\delew^{-1} \sim 3$\%. The light-shaded (green) region is where the
thermal higgsino relic density is smaller than its measured value, with
the balance being made up by something else.  The dashed line shows the
LHC14 reach via the canonical search for gluinos, while the dot-dashed
line shows our projection via searches in the novel SSdB channel
discussed in Sec.~\ref{subsec:lhc}. We see that LHC searches will, by
themselves, not be able to cover the entire parameter space with $\delew
< 30$. The remainder of this parameter space should be accessible, via a
search for higgsinos at an $e^+e^-$ collider operating at
$\sqrt{s}=600$~GeV.

To sum up, we stress that the fact that low scale physics is only
logarithmically (and not quadratically) sensitive to the scale of
ultra-violet physics remains a very attractive feature of softly broken
SUSY models. The fact that it is possible to find phenomenologically
viable models with low $\delew$ leads us to speculate that our
understanding of UV physics is incomplete, and that there might be HS
models with the necessary parameter correlations that will lead to
comparably low values of the true fine-tuning parameter $\delbg$. The
SUSY GUT paradigm remains very attractive despite the absence of new
physics signals at LHC8. We hope that this situation will dramatically
change with the upcoming run of the LHC.

\section*{Acknowledgments}

We are grateful to H.~Baer, V.~Barger, D.~Mickelson, P.~Huang and
W.~Sreethawong for discussions and collaboration on much of the work
described here. We also thank the first three for permitting us to use
of their files for making Fig.~1 in this paper. We thank J.~Kumar for
discussions about fine-tuning, and M.~Drees for comments on the
manuscript.  This work was supported in part by a grant from the US
Department of Energy.


\end{document}